\documentclass[a4paper,11pt]{article}
\usepackage{pos}

\title{Generating airshower images for the VERITAS telescopes with conditional Generative Adversarial Networks}
 \ShortTitle{IACT cGANs}

\author*[a,b]{J. Hoang}
\author[a]{D. A. Williams}
\onbehalf{for the VERITAS Collaboration}

\affiliation[a]{Santa Cruz Institute for Particle Physics,\\
Natural Sciences 2 Building, 1156 High Street, Santa Cruz, CA 95064}

\affiliation[b]{Berkeley SETI Research Center,\\
Campbell Hall, University Drive, Berkeley, CA 94720 - 3411}

\emailAdd{jokhoang@ucsc.edu}
\emailAdd{daw@ucsc.edu}

\abstract{VERITAS (Very Energetic Radiation Imaging Telescope Array System) is the current-generation array comprising four 12-meter optical ground-based Imaging Atmospheric Cherenkov Telescopes (IACTs). Its primary goal is to indirectly observe gamma-ray emissions from the most violent astrophysical sources in the universe. Recent advancements in Machine Learning (ML) have sparked interest in utilizing neural networks (NNs) to directly infer properties from IACT images. However, the current training data for these NNs is generated through computationally expensive Monte Carlo (MC) simulation methods. This study presents a simulation method that employs conditional Generative Adversarial Networks (cGANs) to synthesize additional VERITAS data to facilitate training future NNs. In this test-of-concept study, we condition the GANs on five classes of simulated camera images consisting of circular muon showers and gamma-ray shower images in the first, second, third, and fourth quadrants of the camera. Our results demonstrate that by casting training data as time series, cGANs can 1) replicate shower morphologies based on the input class vectors and 2) generalize additional signals through interpolation in both the class and latent spaces. Leveraging GPUs strength, our method can synthesize novel signals at an impressive speed, generating over $10^6$ shower events in less than a minute.}

\ConferenceLogo{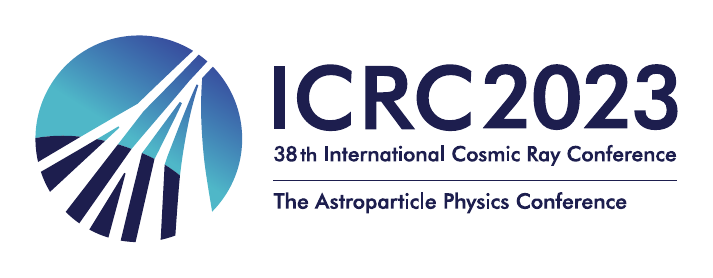}

\FullConference{%
38th International Cosmic Ray Conference (ICRC2023)\\
  26 July - 3 August, 2023\\
  Nagoya, Japan}

\begin{document}
\maketitle

\section{Introduction}
VERITAS is a ground-based gamma-ray instrument located at the Fred Lawrence Whipple Observatory in Arizona \cite{Weekes:2001pd}. Consisting of four IACTs, VERITAS is designed to image brief faint Cherenkov airshowers induced by gamma rays and hadronic cosmic rays. The ambitious next generation of IACTs, the Cherenkov Telescope Array (CTA) \cite{Hofmann:2023fsn} is currently under development. With more advanced instrumentation such as larger mirrors, more sensitive photosensors and upgraded electronics, CTA is expected to improve the current sensitivity by an order of magnitude. There has been an increasing interest in going beyond the classical analysis method employed by the current-generation IACTs toward the state-of-the-art methods based on neural networks (NNs) to reconstruct the properties of each gamma-ray event directly from the telescope images \cite{2021arXiv210514927J}. However, despite the advancements in both hardware and software, an important component in the technique remains largely unchanged: the production of simulated data sets. IACTs’ Monte Carlo (MC) simulation method relies on the use of the CORSIKA software \cite{CORSIKA} to simulate the particle physics of the Extensive Air Showers (EAS) and the Cherenkov light they produce. THe physics-based simulation for IACT involves simulating the first interaction between the gamma ray and the atmosphere, as well as the ray tracing of the Cherenkov photons through the telescopes' optics. As each EAS cascade can produce more than ten thousand Cherenkov photons, the process is a computationally demanding endeavor. 

We demonstrate in this work the feasibility of using a generative machine learning technique called conditional Generative Adversarial Networks (cGANs) to rapidly generate gamma and muon signals for the VERITAS telescopes. cGANs can learn the statistical distribution underlying the training data set, and subsequently generates new images that morphologically resemble the training set. Although the learning process can take several hours, synthesizing new images with cGANs is significantly faster than the traditional physics-based simulation method: the "learned model" is encoded by the ML algorithm as a large matrix; producing new data is thus simply equivalent to a physics-agnostic matrix multiplication process using dedicated GPUs.

\section{Conditional GANs}
GANs \cite{GAN} are a popular framework of generative ML designed to generate images. Each GANs consists of two competing NNs: a Generator and a Discriminator. The Generator is a neural network that takes a random vector \textit{z} from the distribution $p(z)\sim\mathcal{N}(0,1)$ as input and tries to map it to an output similar to the training data. The Discriminator is a classifier that tries to classify whether its inputs come from training data or not. The Generator and Discriminator are trained together in an adversarial game until equilibrium is established.

It is well known that GANs are particularly difficult to train. \textit{Mode collapse} is a commonly encountered problem, occurring when the Generator repeatedly produces a plausible output, and the Discriminator only learns to always reject that plausible output over and over again. To gain more direct control over what GANs should generate, a variant of GANs called conditional Generative Adversarial Networks (cGANs) \cite{cGAN} was invented, feeding a label vector \textit{c} into both the Generator and Discriminator. For example, \textit{c} = [1 0 0 0 0] represents the label for muon showers, \textit{c} = [0 1 0 0 0] is associated with gamma-ray showers having a particular feature, and so on. 

\section{Setup and training}
\subsection{Data}
Our training data set is generated using the open source \texttt{ctapipe} software \cite{ctapipe}. It is a python-based software capable of synthesizing shower-like images with VERITAS' (and several other IACTs') camera geometry, displaying camera images of these showers, and fitting shower parameters such as Hillas moments (for gamma-ray) or radius and center (for muon). For the training data set used in this work, we generate five separated classes: muon shower and gamma-ray shower images in the first, second, third, and fourth quadrants of the camera. Their class labels are [1 0 0 0 0], [0 1 0 0 0], [0 0 1 0 0], [0 0 0 1 0], [0 0 0 0 1], respectively. Each class consists of 20,000 showers.  

\begin{figure}[!ht]
\centering
\includegraphics[width=0.90\linewidth]{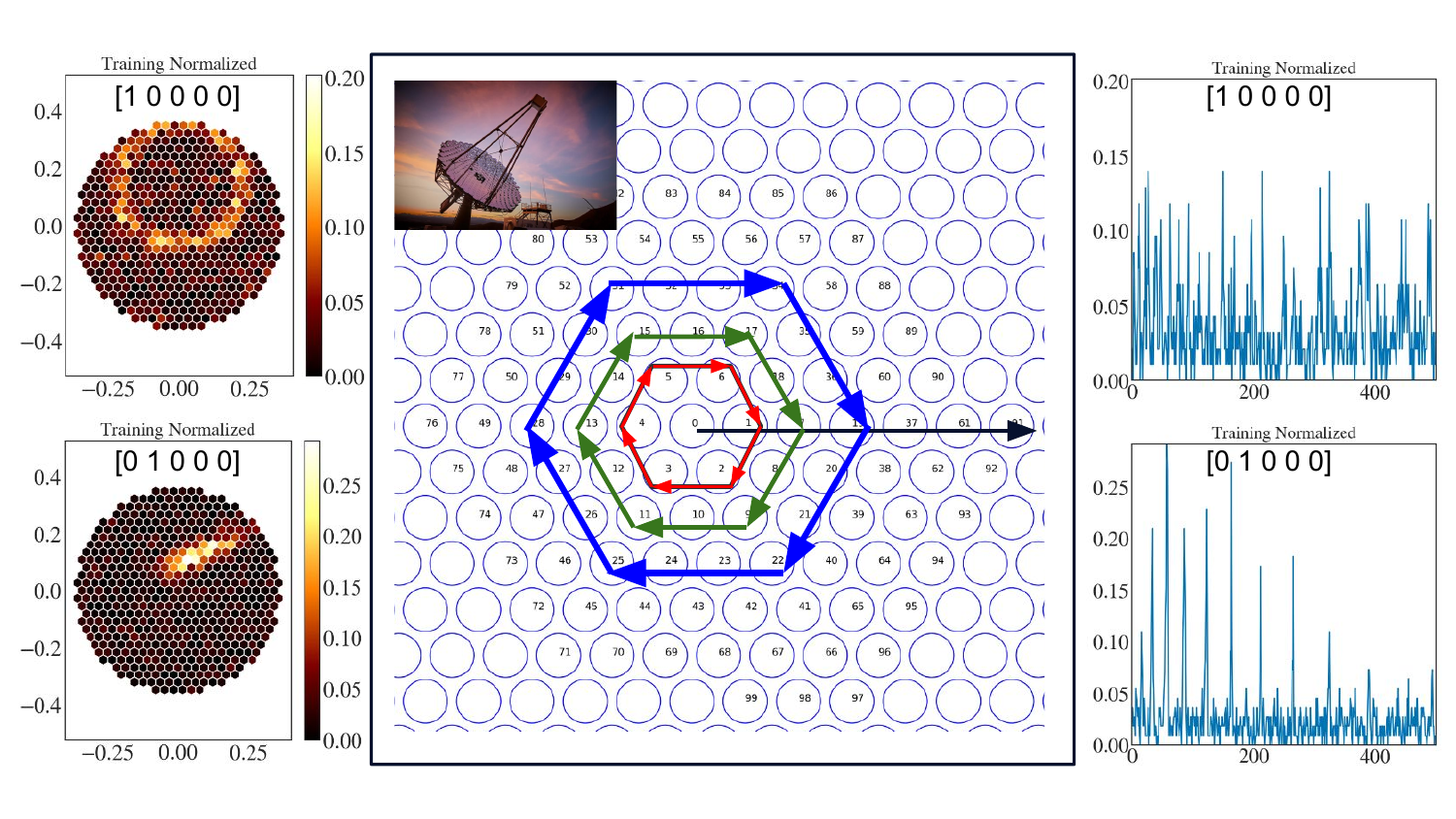} 
\caption{Two different representations of our data. \textbf{Left}: camera view representation reflecting the airshower physics. It is used for visual inspection and shower parameters fitting. \textbf{Middle}: VERITAS's pixel numbering scheme, which follows a concentric hexagonal pattern starting from the central pixel and expanding radially after each revolution. \textbf{Right}: time series representation, used as the Generator's output and the Discriminator's input. The pixel numbering scheme allows us to convert one representation into the other.}
\label{fig:Pixel_mapping}
\end{figure}

Unlike the conventional image-based method, here we work directly at the camera pixel level. In other words, each shower image is cast as a time series with time represented by the pixel index \textit{i} and intensity scaled with the number of photonelectrons detected by the $i^{th}$ photosensor. The image-based camera view representation will be used for visual inspection and fitting shower parameters, while the time series representation will be used for cGANs data input/output. See Fig.\ref{fig:Pixel_mapping} for further explanation.

We choose the abstract time series representation during training for several practical reasons. Firstly, since VERITAS' 499 cylindrical PMT photosensors are arranged in a hexagonal grid, the coordinate transformation of the image into a Cartesian grid will introduce a conversion bias. Secondly, in order to achieve a reasonable image resolution, the required number of square pixels per one hexagonal tile should be greater than one. This leads to an increase in data volume, hence longer training time. Thirdly, even more pixels are required to pad the corners to make a square image. Finally, the time series representation offers a natural pathway to include the label (class) vectors into the training data, since the label vectors described earlier can also be viewed as time series. 

\subsection{NNs architecture and training}
For training and synthesizing new GANs data, we use the TensorFlow machine learning python platform \cite{TensorFlow}. Before training, we normalize the training data so that the range of the input is in the interval [0,1]. Our Discriminator consists of several fully-connected layers of feedforward neural network to gradually \textit{downsample} the concatenated (499+5)-D input vector to 512 $\rightarrow$ 256 $\rightarrow$ 128 $\rightarrow$ 64 $\rightarrow$ 32 $\rightarrow$ 1 output. We also include a layer of Dropout and LeakyReLU after each downsampling layer to prevent overfitting and add a degree of non-linearity to the Discriminator. Similarly, after the concatenation layer to merge the latent and class vectors, our Generator consists of multiple layers that \textit{upsample} the combined (100+5)-D latent and class vector inputs to 128 $\rightarrow$ 256 $\rightarrow$ 512 $\rightarrow$ 1024 $\rightarrow$ 499 dimensional output. We add a layer of BatchNorm to stabilize the training and LeakyReLU for non-linearity after each upsampling step. Overall, the Generator has 1,218,803 trainable parameters, and the Discriminator has 433,153 trainable parameters. 

\begin{figure}[!ht]
\centering
\includegraphics[width=0.75\linewidth]{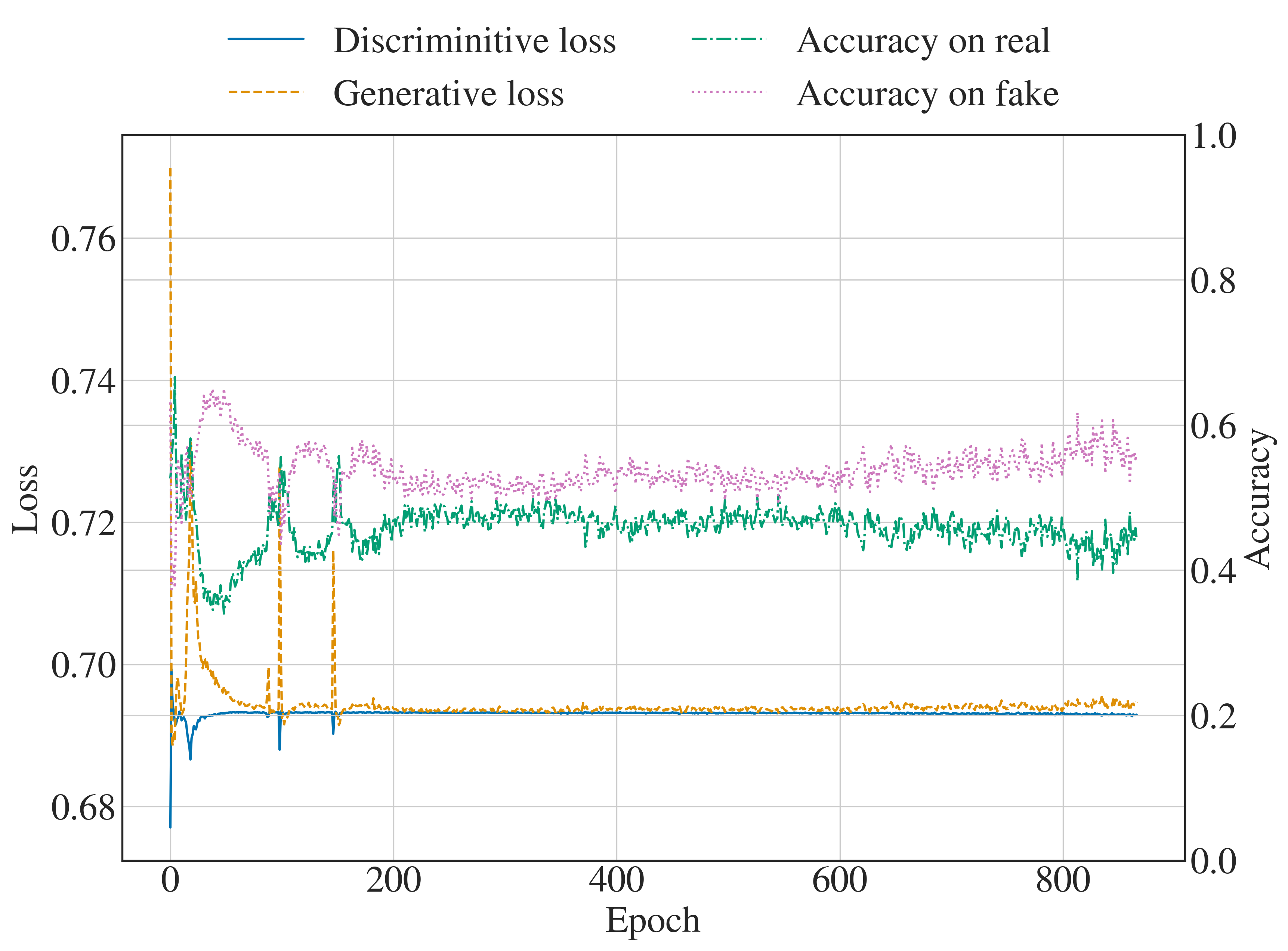} 
\caption{Loss plot during training. Notice how the Discriminator's accuracy score settles around 50\% for real (training) and 50\% for fake (synthesized) data.}
\label{fig:loss_plot}
\end{figure}

In terms of hardware specifications, we performed our training on an M1 Ultra Mac Studio with a configuration of 20-core CPU, 48-core GPU, and 64 GB of unified memory shared between CPU and GPU. With a batch size of 512, it took approximately 10 minutes to go through 20 epochs of training. At the end of every 20 epochs, the program saved the Generator and Discriminator models as .h5 files so that we could subsequently evaluate their outputs. Their file volumes are 4.9 and 5.3 MB, respectively. Currently, there is no notion of a definitive stopping epoch, so we monitor the loss plot (Fig.\ref{fig:loss_plot}) and visually inspect the Generator's outputs to determine when training should be stopped.

\section{Results}

\subsection{Known class reproduction}
\begin{figure}[!ht]
\centering
\includegraphics[width=0.75\linewidth]{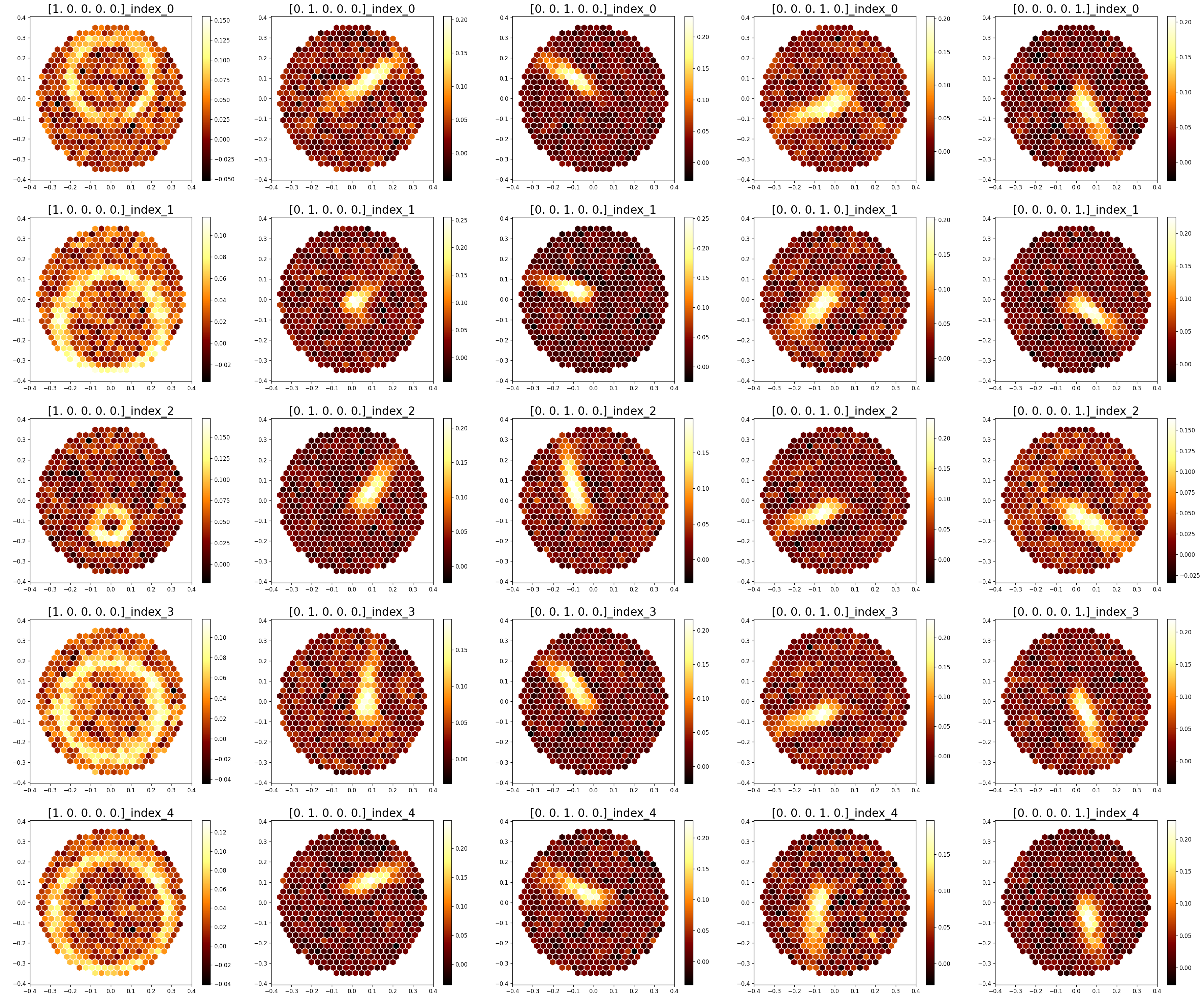} 
\caption{Random cGANs' generated airshowers associated with their class vectors. The zeroth column consists of circular muon showers, first column consists of showers contained in the first quadrant of the camera, etc.}
\label{fig:known_class}
\end{figure}

Fig.\ref{fig:known_class} shows an example of Generator's output at epoch 240. The plot shows a wide variety of signals within each class (column), demonstrating that we have successfully avoided \textit{mode collapse} and generated new signals. Between the classes, the morphologies of signals differ significantly from each other, indicating that the model has learned to associate a typical morphology with a particular label vector.

\begin{figure}[!ht]
\centering
\includegraphics[width=0.99\linewidth]{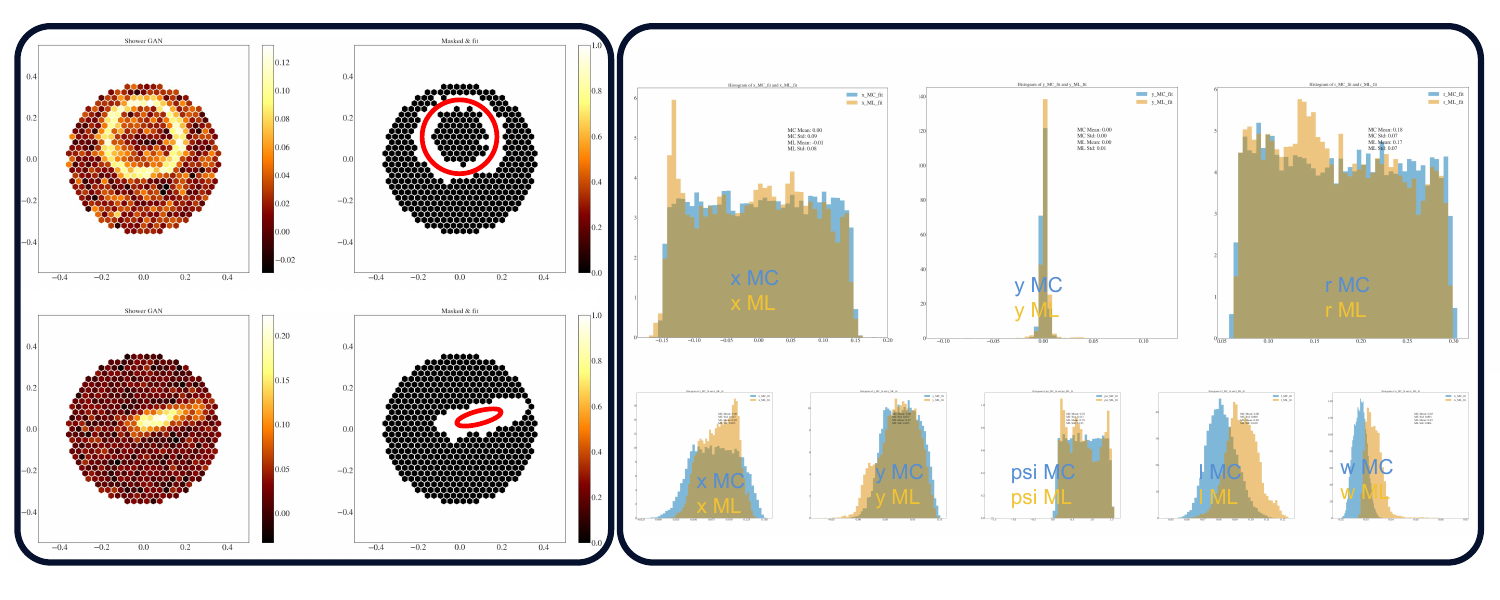} 
\caption{Benchmarking cGANs' outputs with our original training data. \textbf{Left}: Camera view of muon and gamma-ray showers and the masks used for fitting shower parameters. \textbf{Right}: Parameter distributions for muon (center coordinates and radius) and gamma (center coordinates, angle of the major axis with respect to the x-axis, length, and width).}
\label{fig:bench_mark}
\end{figure}

To benchmark how similar the Generator's synthesized data are to the training data set, we ask the Generator to synthesize a population of 20,000 showers for each class, use \texttt{ctapipe} to fit basic parameters from the camera images, and compare the distributions of the parameters with the training data set. The results for class 0 and 1 are shown in Fig.\ref{fig:bench_mark}. Class 0 (muon) has similar distributions between training and GAN-synthesized data, except for edge cases where there are insufficient training data. Class 1 (gamma rays in the first quadrant) shows similar means and standard deviations for the center and angle distributions, but there is a notable systematic difference in the width and length distributions. This could be because either the training data set was not sufficient, bias in fitting methods, or poor network architecture.

\subsection{Interpolation of class and latent space vectors}
\begin{figure}[!ht]
\centering
\includegraphics[width=0.99\linewidth]{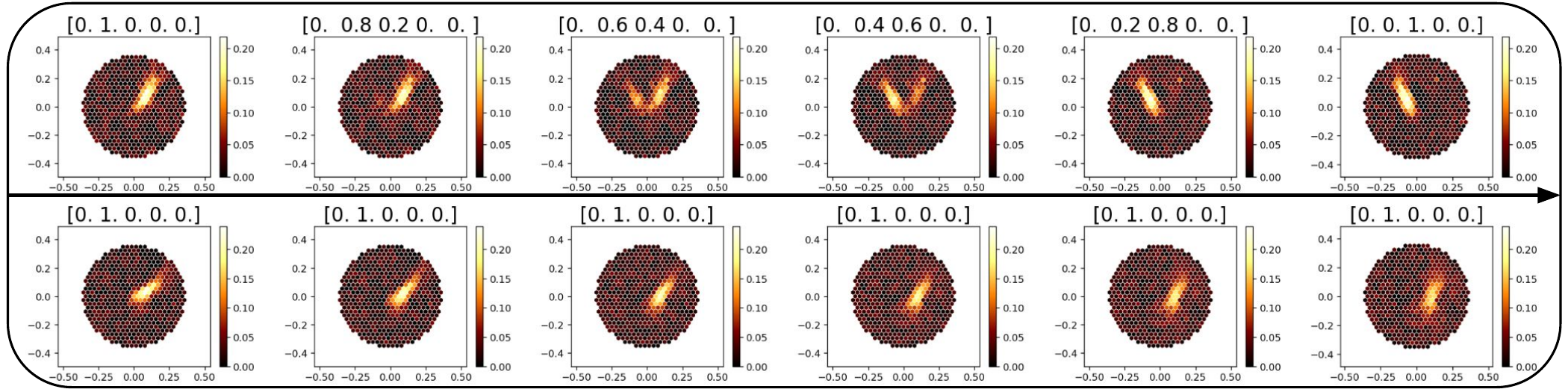} 
\caption{Differences between class and vector space interpolation. \textbf{Top}: Physical interpretation when we perform a linear walk from class 1 to class 2 while remaining stationary in the latent space. \textbf{Bottom}: Physical interpretation when we perform a linear walk between two random points in the latent space while keeping the class vector fixed.}
\label{fig:Interpolate}
\end{figure}

To probe how the Generator interprets the input vectors, we perform two experiments. In the first experiment, we interpolate between two points in the class space while keeping the latent vector fixed. In the second experiment, we choose a fixed class vector, select two randomly generated latent vectors and linearly interpolate the steps between them. An example is shown in Fig.\ref{fig:Interpolate}

With the first walk in class space, our cGANs could not have known about the existence of intermediate vectors between two class spaces: during training, every class vector label consists of only 0s and 1 entries. Therefore, it is apparent that as we move toward the midpoint between the two classes, the two middle columns show non-physical results. The latent space walk, however, is different. Since we have sampled many latent vectors from the latent space during training, our cGANs would have known about the existence of intermediate vectors populating this space. As such, the transition appears to be smoother and more physically plausible. 

\section{Discussion}

\begin{figure}[!ht]
\centering
\includegraphics[width=0.80\linewidth]{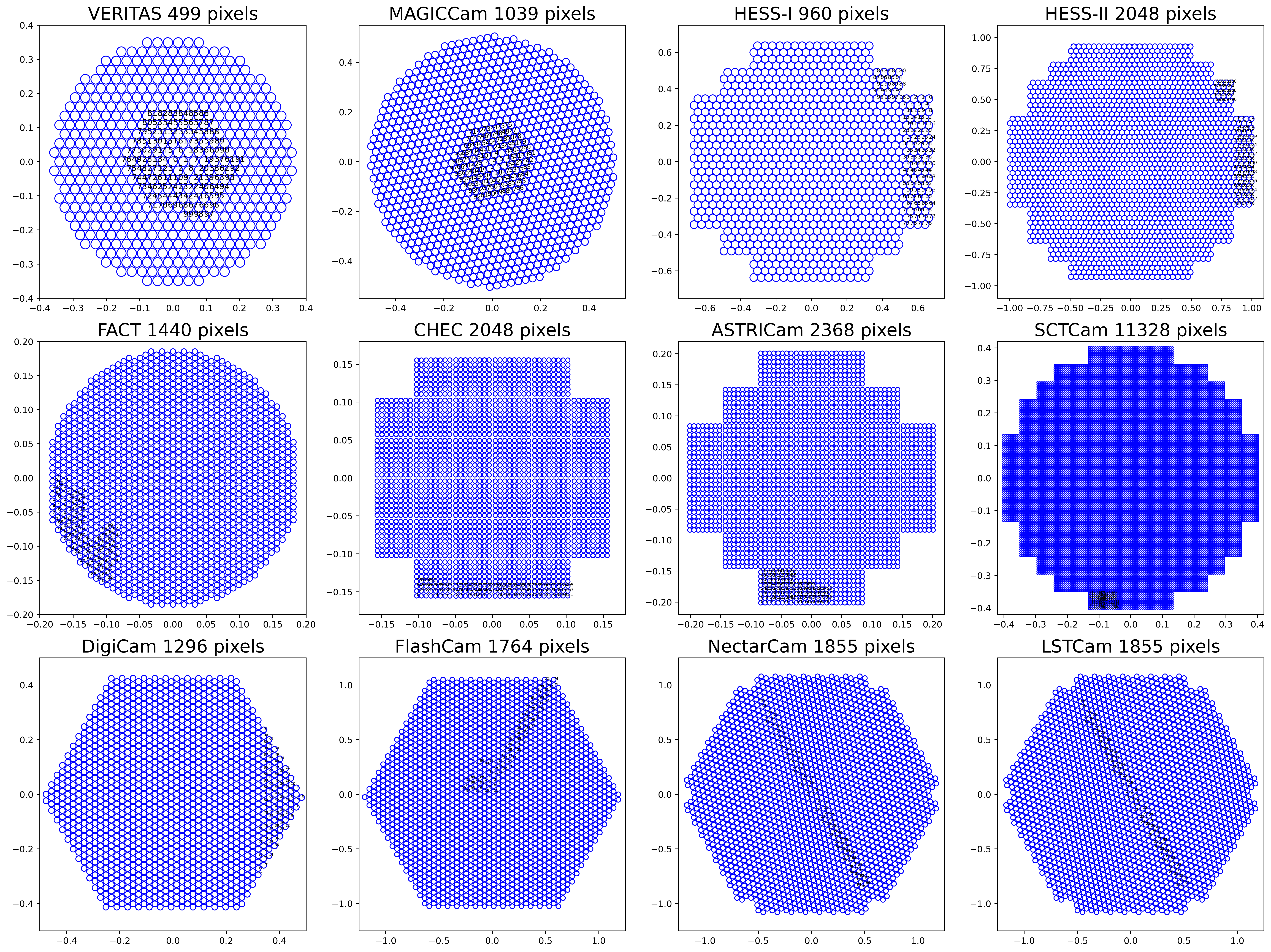} 
\caption{Position and index of the first 100 pixels from some of the current and next-generation IACT cameras.}
\label{fig:IACT_cameras}
\end{figure}

We highlight the interesting discovery that cGANs managed to learn the underlying statistical distribution of the time series training data. Recall that we have provided an implicit projection by going from the 2D camera view representation into the 1D time series representation; such mapping between the two representations is definitely non-trivial. Therefore it is surprising that our Generator can produce consistent signals with complex morphologies such as circular muon rings without relying on this mapping.

Furthermore, we note that VERITAS's pixel arrangement is not unique among Cherenkov telescopes: newer-generation telescopes such as the prototype SCT (pSCT) employ square SiPMs arranged in a Cartesian grid. We have also performed further experiments on some types of IACT camera geometry shown in Fig.\ref{fig:IACT_cameras}, and preliminary results suggest the general applicability of the method. Therefore, we conjecture that cGANs can reproduce consistent airshower images for any IACT camera regardless of the arbitrary “hidden” pixel mapping, and encourage further exploration on the topic. 

Finally, the relatively small file size of the Generator's model offers two potential applications. It can be employed as a data compression scheme and a data expansion tool. After abstracting the training data into an NN-based model consisting of a few million parameters, one can potentially generate on the fly as much data as needed from any local machine using only TensorFlow. Another possibility is just-in-time simulation, which eliminates the need to store and transfer a large corpus of data across multiple geographical locations, freeing up computing overheads. In our setup, the trained Generator manages to generate $10^{6}$ showers in $\sim$70 seconds using a single terminal. However, the machine is also capable of running four such processes in parallel at a reduced speed, i.e. approximately 90 seconds per $10^{6}$ showers. In other words, more than a million showers can be generated in under a minute.

\section{Conclusion}
This work demonstrates the successful avoidance of mode collapse by our cGANs and the generation of a novel set of VERITAS data. The majority of the generated events exhibit morphological similarities to those generated using the Monte Carlo method. Additionally, the speed of event generation using cGANs is unprecedented. We anticipate further enhancements in the accuracy of the outputs as we continue to fine-tune our experimental setup and NN architecture.

\section{Acknowledgments}
We would like to express our gratitude to the Breakthrough Listen Initiative and UC Berkeley for their generous support. Special thanks to D. Ribeiro, D. Nieto, T. Hassan, and S. Spencer for their valuable contributions and insightful discussions. This work is funded by the BL Initiative and the NSF.

\bibliographystyle{JHEP}
\bibliography{refs}

%
%
%

\clearpage

\section*{Full Author List: VERITAS Collaboration}

\scriptsize
\noindent
A.~Acharyya$^{1}$,
C.~B.~Adams$^{2}$,
A.~Archer$^{3}$,
P.~Bangale$^{4}$,
J.~T.~Bartkoske$^{5}$,
P.~Batista$^{6}$,
W.~Benbow$^{7}$,
J.~L.~Christiansen$^{8}$,
A.~J.~Chromey$^{7}$,
A.~Duerr$^{5}$,
M.~Errando$^{9}$,
Q.~Feng$^{7}$,
G.~M.~Foote$^{4}$,
L.~Fortson$^{10}$,
A.~Furniss$^{11, 12}$,
W.~Hanlon$^{7}$,
O.~Hervet$^{12}$,
C.~E.~Hinrichs$^{7,13}$,
J.~Hoang$^{12}$,
J.~Holder$^{4}$,
Z.~Hughes$^{9}$,
T.~B.~Humensky$^{14,15}$,
W.~Jin$^{1}$,
M.~N.~Johnson$^{12}$,
M.~Kertzman$^{3}$,
M.~Kherlakian$^{6}$,
D.~Kieda$^{5}$,
T.~K.~Kleiner$^{6}$,
N.~Korzoun$^{4}$,
S.~Kumar$^{14}$,
M.~J.~Lang$^{16}$,
M.~Lundy$^{17}$,
G.~Maier$^{6}$,
C.~E~McGrath$^{18}$,
M.~J.~Millard$^{19}$,
C.~L.~Mooney$^{4}$,
P.~Moriarty$^{16}$,
R.~Mukherjee$^{20}$,
S.~O'Brien$^{17,21}$,
R.~A.~Ong$^{22}$,
N.~Park$^{23}$,
C.~Poggemann$^{8}$,
M.~Pohl$^{24,6}$,
E.~Pueschel$^{6}$,
J.~Quinn$^{18}$,
P.~L.~Rabinowitz$^{9}$,
K.~Ragan$^{17}$,
P.~T.~Reynolds$^{25}$,
D.~Ribeiro$^{10}$,
E.~Roache$^{7}$,
J.~L.~Ryan$^{22}$,
I.~Sadeh$^{6}$,
L.~Saha$^{7}$,
M.~Santander$^{1}$,
G.~H.~Sembroski$^{26}$,
R.~Shang$^{20}$,
M.~Splettstoesser$^{12}$,
A.~K.~Talluri$^{10}$,
J.~V.~Tucci$^{27}$,
V.~V.~Vassiliev$^{22}$,
A.~Weinstein$^{28}$,
D.~A.~Williams$^{12}$,
S.~L.~Wong$^{17}$,
and
J.~Woo$^{29}$\\
\\
\noindent
$^{1}${Department of Physics and Astronomy, University of Alabama, Tuscaloosa, AL 35487, USA}

\noindent
$^{2}${Physics Department, Columbia University, New York, NY 10027, USA}

\noindent
$^{3}${Department of Physics and Astronomy, DePauw University, Greencastle, IN 46135-0037, USA}

\noindent
$^{4}${Department of Physics and Astronomy and the Bartol Research Institute, University of Delaware, Newark, DE 19716, USA}

\noindent
$^{5}${Department of Physics and Astronomy, University of Utah, Salt Lake City, UT 84112, USA}

\noindent
$^{6}${DESY, Platanenallee 6, 15738 Zeuthen, Germany}

\noindent
$^{7}${Center for Astrophysics $|$ Harvard \& Smithsonian, Cambridge, MA 02138, USA}

\noindent
$^{8}${Physics Department, California Polytechnic State University, San Luis Obispo, CA 94307, USA}

\noindent
$^{9}${Department of Physics, Washington University, St. Louis, MO 63130, USA}

\noindent
$^{10}${School of Physics and Astronomy, University of Minnesota, Minneapolis, MN 55455, USA}

\noindent
$^{11}${Department of Physics, California State University - East Bay, Hayward, CA 94542, USA}

\noindent
$^{12}${Santa Cruz Institute for Particle Physics and Department of Physics, University of California, Santa Cruz, CA 95064, USA}

\noindent
$^{13}${Department of Physics and Astronomy, Dartmouth College, 6127 Wilder Laboratory, Hanover, NH 03755 USA}

\noindent
$^{14}${Department of Physics, University of Maryland, College Park, MD, USA }

\noindent
$^{15}${NASA GSFC, Greenbelt, MD 20771, USA}

\noindent
$^{16}${School of Natural Sciences, University of Galway, University Road, Galway, H91 TK33, Ireland}

\noindent
$^{17}${Physics Department, McGill University, Montreal, QC H3A 2T8, Canada}

\noindent
$^{18}${School of Physics, University College Dublin, Belfield, Dublin 4, Ireland}

\noindent
$^{19}${Department of Physics and Astronomy, University of Iowa, Van Allen Hall, Iowa City, IA 52242, USA}

\noindent
$^{20}${Department of Physics and Astronomy, Barnard College, Columbia University, NY 10027, USA}

\noindent
$^{21}${ Arthur B. McDonald Canadian Astroparticle Physics Research Institute, 64 Bader Lane, Queen's University, Kingston, ON Canada, K7L 3N6}

\noindent
$^{22}${Department of Physics and Astronomy, University of California, Los Angeles, CA 90095, USA}

\noindent
$^{23}${Department of Physics, Engineering Physics and Astronomy, Queen's University, Kingston, ON K7L 3N6, Canada}

\noindent
$^{24}${Institute of Physics and Astronomy, University of Potsdam, 14476 Potsdam-Golm, Germany}

\noindent
$^{25}${Department of Physical Sciences, Munster Technological University, Bishopstown, Cork, T12 P928, Ireland}

\noindent
$^{26}${Department of Physics and Astronomy, Purdue University, West Lafayette, IN 47907, USA}

\noindent
$^{27}${Department of Physics, Indiana University-Purdue University Indianapolis, Indianapolis, IN 46202, USA}

\noindent
$^{28}${Department of Physics and Astronomy, Iowa State University, Ames, IA 50011, USA}

\noindent
$^{29}${Columbia Astrophysics Laboratory, Columbia University, New York, NY 10027, USA}

\end{document}